\documentclass[reprint,twocolumn,aps,prb,showpacs]{revtex4-1}

\usepackage{amsmath}
\usepackage{amssymb}
\usepackage{graphicx}
\usepackage{dcolumn}
\usepackage{bm}
\usepackage[colorlinks=true, allcolors=blue]{hyperref}
\usepackage{epstopdf}
\usepackage[utf8]{inputenc}

\begin{document}

\title{Impurity-Scattering Assisted Umklapp Scattering as the Origin of
Low-Temperature Resistivity in the Normal-State of Cuprate Superconductors}

\author{Xingyu Ma$^{1}$}
\author{Minghuan$^{2}$}
\email{mhzeng@cqu.edu.cn}
\author{Huaiming Guo$^{3}$}
\author{Shiping Feng$^{4}$}
\email{spfeng@bnu.edu.cn}

\affiliation{$^{1}$Institute of Physics, Chinese Academy of Sciences, Beijing 
100190, China}

\affiliation{$^{2}$College of Physics, Chongqing University, Chongqing 401331, 
China}

\affiliation{$^{3}$School of Physics, Beihang University, Beijing 100191, 
China}

\affiliation{$^{4}$School of Physics and Astronomy, Beijing Normal University, 
Beijing 100875, China and \\
Department of Physics, Beijing Normal University, Zhuhai 519087, China}


\begin{abstract}
The transport experiments reveal that the low-temperature resistivity in the
normal-state of cuprate superconductors is quadratic in temperature (T-quadratic) in
the underdoped pseudogap phase, while it is linear in temperature (T-linear) in the
overdoped strange-metal phase, however, the full understanding of these different
behaviours is still a challenging issue. Here starting from the microscopic
electronic structure of cuprate superconductors, the low-temperature resistivity in
the normal-state is investigated from the underdoped pseudogap phase to the overdoped
strange-metal phase. It is shown that the mechanism requires both the impurity
scattering and the umklapp scattering: the impurity scattering is needed to restrict
the modification of the distribution function to at around the antinodal region,
while the impurity-scattering assisted umklapp scattering from a spin excitation is
at the heart of the behaviour in the low-temperature resistivity, where the doping
dependence of the temperature scale exists, and presents a similar behavior of the
antinodal spin pseudogap crossover temperature. In the low-temperature region above
the temperature scale in the overdoped strange-metal phase, the resistivity is
T-linear, however, in the low-temperature region below the temperature scale in the
underdoped pseudogap phase, the opening of the spin pseudogap lowers the spin
excitation density of states at around the antinodal region, which reduces the
strength of the electron umklapp scattering from a spin excitation associated with
the antinode, and thus leads to a T-quadratic behaviour of the resistivity.
\end{abstract}


\maketitle

\section{Introduction}\label{Introduction}

The parent compound of cuprate superconductors is a Mott insulator
\cite{Kastner98,Fujita12} with an antiferromagnetic (AF) long-range order (AFLRO),
however, after this AFLRO is destroyed rapidly by the charge-carrier doping,
superconductivity emerges \cite{Bednorz86}. In addition to their unique
superconducting (SC) properties \cite{Damascelli03,Campuzano04,Fink07}, the cuprate
superconductors have a rich temperature-doping phase diagram
\cite{Damascelli03,Campuzano04,Fink07,Drozdov18}, where three distinct regions need
to be distinguished: (a) the optimally doped regime, where the highest SC transition
temperature $T_{\rm c}$ occurs \cite{Drozdov18}; (b) the underdoped regime, where an
energy gap \cite{Timusk99,Hufner08,Hussey11,Vishik18} called the normal-state
pseudogap exists above $T_{\rm c}$ but below the pseudogap crossover temperature
$T^{*}$. This is why in the underdoped regime, the phase above $T_{\rm c}$ but below
$T^{*}$ has been refereed to as {\it the pseudogap phase}. In particular, this
normal-state pseudogap is present in both the spin and charge channels
\cite{Timusk99,Hufner08,Hussey11,Vishik18,Loeser96,Norman98,Renner98,Warren89,Alloul89,Walstedt90},
and then the physical response in the underdoped regime can be well interpreted in
terms of the opening of the normal-state pseudogap; (c) the overdoped regime, where
the normal-state in low temperatures exhibits a number of the anomalous properties
\cite{Keimer15,Varma20,Hussey23} in the sense that they do not fit in with the
conventional Fermi-liquid theory \cite{Schrieffer64,Abrikosov88,Mahan81}, which has led
to the normal-state in the overdoped regime being refereed to as
{\it the strange-metal phase}.

The understanding of the nature of the normal-state in cuprate superconductors
represents a formidable challenge for theory \cite{Keimer15,Varma20,Hussey23}, since
superconductivity itself is a corollary to the normal state properties. One essential
ingredient in the quest for the nature of the normal-state is the resistivity
\cite{Keimer15,Varma20,Hussey23}. In this paper, we shall not address the case at
high temperature, but instead focus on the low-temperature resistivity, since whose
origin is intertwined with the mechanism of superconductivity
\cite{Keimer15,Varma20,Hussey23}. Experimentally, by virtue of systematic studies
using the transport measurement technique, the low-temperature resistivity from the
underdoped pseudogap phase to the overdoped strange-metal phase is well-established
by now, where some agreements have emerged that (a) the variation of the
low-temperature resistivity is quadratic in temperature (T-quadratic) for a wide
doping range in the underdoped pseudogap phase
\cite{Bucher93,Ito93,Nakano94,Ando01,Cooper09,Mirzaei13,Barisic13,Pelc20};
(b) the low-temperature resistivity increases linearly with temperature in the
overdoped strange-metal phase extending up to the edge of the SC dome
\cite{Legros19,Ayres21,Grisso21,Gurvitch87}; and (c) the strengths of both the
T-quadratic resistivity and the linear in temperature (T-linear) resistivity decrease
with the increase of doping. These transport experimental observations therefore
indicate an exceptional crossover from the low-temperature T-quadratic resistivity
in the underdoped pseudogap phase to the low-temperature T-linear resistivity in the
overdoped strange-metal phase, however, a complete understanding of this exotic
crossover is still unclear. On the other hand, the low-temperature T-linear
resistivity and the related strange-metal behaviours have been observed recently in
artificial cuprate superlattices and quantum-engineered heterostructures
\cite{Campi24,Campi25}, where the interplay between the Fermi-liquid behavior,
strange-metal behavior, and superconductivity can be tuned via geometric and
band-structure control. These experimental results \cite{Campi24,Campi25} therefore
offer a complementary perspective for a controlled crossover between the Fermi-liquid
and strange-metal behaviors in a doped Mott insulator.

Theoretically, several mechanisms have been suggested for the interpretation of the
origins of the low-temperature T-quadratic resistivity in the underdoped pseudogap
phase and the low-temperature T-linear resistivity in the overdoped strange-metal
phase, where the T-linear resistivity in the overdoped strange-metal phase was
explained as a consequence of the scale invariant physics near to the quantum
critical point \cite{Damle97,Sachdev11}, or from the Planckian dissipation
\cite{Zaanen04,Luca07,Zaanen19,Hartnoll22}. In particular, the T-linear resistivity
in the underdoped regime above $T^{*}$ has been attributed to the electron umklapp
scattering \cite{Rice17}, however, when the temperatures fall below $T^{*}$, the
opening of the pseudogap restricts the available umklapp scattering channels, which
generates T-quadratic resistivity in the underdoped pseudogap phase at the
temperature below $T^{*}$. The charge-carrier doping process nearly always introduces
some measure of disorder \cite{Hussey02,Balatsky06,Alloul09}, leading to that in
principle, all cuprate superconductors have naturally impurities, and then the
impurity-scattering effect may play an important role both in the normal-state and
the SC-state properties. In this case, it was shown that the low-temperature T-linear
resistivity in the overdoped strange-metal phase is induced by the impurity-scattering
assisted umklapp scattering from a critical boson mode \cite{Lee21}. Moreover, in the
case of {\it near} the clean limit, we \cite{Ma23,Ma24} have also studied recently the
nature of the low-temperature resistivity from the underdoped pseudogap phase to the
overdoped strange-metal phase, where the low-temperature T-linear resistivity in the
overdoped strange-metal phase originates mainly from the impurity-scattering assisted
antinodal umklapp scattering between electrons by the exchange of the effective spin
propagator, however, when this impurity-scattering assisted electron antinodal umklapp
scattering flows to the underdoped pseudogap phase, the opening of the antinodal spin
pseudogap decreases the spin excitation density of states at around the antinodal
region, which reduces the strength of the impurity-scattering assisted electron
antinodal umklapp scattering, and therefore leads to the low-temperature T-quadratic
resistivity. In these previous studies \cite{Ma23,Ma24}, our main goal is to emphasize
that the impurity-scattering assisted electron umklapp scattering from a spin
excitation is at the heart of the behaviour in the low-temperature resistivity, and
therefore the crucial role played by the impurity scattering itself is not discussed
in details. In this paper, as a complement of these previous works \cite{Ma23,Ma24},
we restudy the nature of the low-temperature resistivity from the underdoped
pseudogap phase to the overdoped strange-metal phase, and show that the impurity
scattering is required to restrict the modification of the distribution function to
at around the antinodal region, and then the impurity-scattering assisted electron
umklapp scattering associated with the antinode is responsible for the low-temperature
resistivity in the normal-state of cuprate superconductors.

This paper is organized as follows. In Section \ref{Formalism}, we briefly review the
microscopic theory of the pseudogap in both the spin and charge channels. In
particular, the normal-state pseudogap suppresses partially the spectral weight on the
electron Fermi surface (EFS) at around the antinodal region, and then the important
electron umklapp scattering process is concentrated at around the antinodal region.
Starting from the microscopic electronic structure, the resistivity in the
normal-state due to the
impurity-scattering assisted electron umklapp scattering from a spin excitation is
derived in terms of the Boltzmann transport equation. The quantitative characteristics
of the low-temperature resistivity are presented in Section \ref{electron-resistivity},
where we show that the doping dependence of the temperature scale is proportional to the
square of the minimal electron umklapp scattering vector, and presents a similar
behavior of the antinodal spin pseudogap crossover temperature. In the low-temperature
region above the temperature scale in the overdoped strange-metal phase, the
resistivity exhibits a T-linear behaviour, while in the low-temperature region below
the temperature scale in the underdoped pseudogap phase, where antinodal spin pseudogap
suppresses partially the spin excitation density of states at around the antinodal
region, which decreases the strength of the impurity-scattering assisted electron
antinodal umklapp scattering, and thus generates a T-quadratic behaviour of the
resistivity. Finally, we give a summary in Section \ref{summary}. In the
Appendix \ref{NSBE}, we present the detailed form for the numerical solution of the
Boltzmann equation, and show that the low-temperature resistivity in the clean limit
shrinks to zero.

\section{Theoretical Framework}\label{Formalism}

The basic element in the layered crystal structure of cuprate superconductors is the
square-lattice copper-oxide layer \cite{Bednorz86}, and it is widely believed that the
nonconventional features of cuprate superconductors are mainly governed by these
copper-oxide layers. In this case, it has been proposed that the fundamental physics
of the doped copper-oxide layer is contained in the low-energy effective $t$-$J$ model
on a square lattice \cite{Anderson87},
\begin{eqnarray}\label{tJ-model}
H &=&-t\sum_{l\hat{\eta}\sigma}C^{\dagger}_{l\sigma}C_{l+\hat{\eta}\sigma}
+t'\sum_{l\hat{\tau}\sigma}C^{\dagger}_{l\sigma}C_{l+\hat{\tau}\sigma}
\nonumber\\
&+&\mu\sum_{l\sigma}C^{\dagger}_{l\sigma}C_{l\sigma}
+J\sum_{l\hat{\eta}}{\bf S}_{l}\cdot {\bf S}_{l+\hat{\eta}},~~~~
\end{eqnarray}
supplemented by the on-site local constraint
$\sum_{\sigma}C_{l\sigma}^{\dagger}C_{l\sigma}\leq 1$ to remove double occupancy of
any a site, where the operator $C^{\dagger}_{l\sigma}$ and operator $C_{l\sigma}$
creates and annihilates an electron with spin $\sigma$ at site $l$, ${\bf S}_{l}$ is
the spin operator with its components $S_{l}^{x}$, $S_{l}^{y}$, and $S_{l}^{z}$, and
$\mu$ is the chemical potential. The summation is taken over all sites $l$, and for
each site $l$, restricted to its nearest-neighbor (NN) sites $\hat{\eta}$ or next NN
sites $\hat{\tau}$. Throughout this paper, the magnetic coupling $J$ and the lattice
constant of the square lattice are set as the energy and length units, respectively,
while the parameters \cite{Ma23,Ma24} in the low-energy effective $t$-$J$ model
(\ref{tJ-model}) are chosen as $t/J=2.5$ and $t'/t=0.3$, which are the typical values
of cuprate superconductors \cite{Damascelli03,Campuzano04,Fink07,Kim98}. Moreover,
when necessary to compare with the experimental data, we set $J=100$meV. However, it
should be noted that in cuprate superconductors, the values of $t$, $t'$, and $J$ are
believed to vary somewhat from compound to compound, and then some differences among
the different families have been observed experimentally
\cite{Damascelli03,Campuzano04,Fink07,Kim98}. In particular, it has been shown
experimentally \cite{Damascelli03,Campuzano04,Fink07,Kim98} and theoretically
\cite{Hybertsen90,Gooding94,Tanaka04,Shih04,Pavarini01} that these differences for
different families of cuprate superconductors are mainly correlated with $t'$. In
this case, we \cite{Liu21,Tan21,Gao18,Feng06} have made a series of calculations for
the low-energy electronic structure and the related $T_{\rm c}$ with different
values of $t'$, and the obtained results are qualitatively consistent with the
corresponding experimental data \cite{Damascelli03,Campuzano04,Fink07,Kim98}.

The on-site local constraint of no double electron occupancy is a direct reflection
of the strong electron correlation \cite{Anderson87,Yu92,Feng93,Zhang93,Lee06} in
the low-energy effective $t$-$J$ model (\ref{tJ-model}), and can be treated properly
in terms of the fermion-spin transformation \cite{Feng9404,Feng15},
\begin{eqnarray}\label{CSSFS}
C_{l\uparrow}=h^{\dagger}_{l\uparrow}S^{-}_{l}, ~~~~
C_{l\downarrow}=h^{\dagger}_{l\downarrow}S^{+}_{l},
\end{eqnarray}
where $S^{+}_{l}$ and $S^{-}_{l}$ are the $U(1)$ gauge invariant spin-raising and
spin-lowering operators, respectively, which carry spin index of the constrained
electron, and thus the collective mode from this spin degree of freedom of the
constrained electron is interpreted as the spin excitation responsible for the spin
dynamics of the system, and $h^{\dagger}_{l\sigma}=e^{i\Phi_{l\sigma}}h^{\dagger}_{l}$
and $h_{l\sigma}=e^{-i\Phi_{l\sigma}}h_{l}$ are the $U(1)$ gauge invariant charge
carrier creation and annihilation operators, respectively, which describe the charge
degree of freedom of the constrained electron together with some effects of spin
configuration rearrangements due to the presence of the doped charge carrier itself,
while the constrained electron as a result of the charge-spin recombination of a
charge carrier and a localized spin is responsible for the electronic properties.

\subsection{Normal-State Pseudogap}\label{Phase-diagram}

In cuprate superconductors, the pseudogap state is particularly obvious in the
underdoped regime \cite{Timusk99,Hufner08,Hussey11,Vishik18}, leading to that all the
exotic features of the normal-state in the underdoped regime are correlated directly
to the opening of the pseudogap. Starting from the fermion-spin theory (\ref{CSSFS})
description of the low-energy effective $t$-$J$ model (\ref{tJ-model}), the
microscopic theory of the pseudogap in both the charge and spin channels has been
developed \cite{Feng12,Li25,Feng15a,Feng16,Kuang15}, where the normal-state pseudogap
originates directly from the constrained electrons in the two-dimensional Fermi sea
scattered by the spin excitations (then the collective mode from the spin degree of
freedom of the constrained electron itself), while the spin pseudogap is generated
directly by the coupling of the spin excitations with the charge
carriers (then the charge degree of freedom of the constrained electron itself). The
work in this paper builds on this microscopic pseudogap theory, and here we sketch the
main formalism and results. In these previous discussions \cite{Li25,Feng15a,Feng16},
the full electron propagator of the fermion-spin theory (\ref{CSSFS}) description of
the low-energy effective $t$-$J$ model (\ref{tJ-model}) has been obtained in terms of
the full charge-spin recombination as,
\begin{eqnarray}\label{EGF}
G({\bf k},\omega)={1\over \omega-\varepsilon_{\bf k}
-\Sigma_{\rm ph}({\bf k},\omega)},
\end{eqnarray}
where $\varepsilon_{\bf k}=-4t\gamma_{\bf k}+4t'\gamma_{\bf k}'+\mu$ is the electron
energy dispersion in the tight-binding approximation, with
$\gamma_{\bf k}=({\rm cos}k_{x}+{\rm cos} k_{y})/2$ and
$\gamma_{\bf k}'={\rm cos}k_{x}{\rm cos}k_{y}$, while the electron self-energy
originates from the interaction between electrons by the exchange of the effective
spin propagator, and can be expressed as,
\begin{eqnarray}\label{ESE}
\Sigma_{\rm ph}({\bf k},i\omega_{n})&=&{1\over N}\sum_{\bf p}{1\over\beta}
\sum_{ip_{m}}G({\bf p}+{\bf k},ip_{m}+i\omega_{n})\nonumber\\
&\times&P^{(0)}({\bf k},{\bf p},ip_{m}),~~~
\end{eqnarray}
with the number of lattice sites $N$, the fermion and boson Matsubara frequencies
$\omega_{n}$ and $p_{m}$, respectively, and the mean-field (MF) effective spin
propagator,
\begin{eqnarray}\label{ESP-1}
P^{(0)}({\bf k},{\bf p},\omega)={1\over N}\sum_{\bf q}
\Lambda^{2}_{{\bf p}+{\bf q}+{\bf k}}\Pi({\bf p},{\bf q},\omega),
\end{eqnarray}
where $\Lambda_{{\bf k}}=4t\gamma_{\bf k}-4t'\gamma_{\bf k}'$ is the vertex function,
and $\Pi({\bf p},{\bf q},\omega)$ is the MF spin bubble, which is a convolution of
two MF spin propagators, and can be expressed explicitly as,
\begin{equation}\label{MF-spin-bubble}
\Pi({\bf p},{\bf q},ip_{m})={1\over\beta}\sum_{iq_{m}}D^{(0)}({\bf q},iq_{m})
D^{(0)}({\bf q}+{\bf p},iq_{m}+ip_{m}),~~~~
\end{equation}
with the boson Matsubara frequency $q_{m}$, and the MF spin propagator,
\begin{eqnarray}\label{SGF-1}
D^{(0)}({\bf k},\omega)={B_{\bf k}\over\omega^{2}-\omega^{2}_{\bf k}}
={B_{\bf k}\over 2\omega_{\bf k}}\left ( {1\over\omega-\omega_{\bf k}}
-{1\over\omega+\omega_{\bf k}}\right ),~~~~~
\end{eqnarray}
where the MF spin excitation energy dispersion $\omega_{\bf k}$ and the corresponding
spectral weight $B_{\bf k}$ have been given in Ref. \onlinecite{Kuang15}.

The key observation in the theory \cite{Feng12,Li25,Feng15a,Feng16} is that the
normal-state pseudogap is closely associated with the electron self-energy as
$\Sigma_{\rm ph}({\bf k},\omega) \approx [2\bar{\Delta}_{\rm PG}({\bf k})]^{2}
/[\omega -\varepsilon_{0{\bf k}}]$, where $\bar{\Delta}_{\rm PG}({\bf k})$ is
identified as being a region of the electron self-energy in which
$\bar{\Delta}_{\rm PG}({\bf k})$ partially suppresses the electronic density of
states on EFS, and in this sense, $\bar{\Delta}_{\rm PG}({\bf k})$ has been
refereed to as the normal-state pseudogap, with the normal-state pseudogap parameter
$\bar{\Delta}^{2}_{\rm PG}=(1/N)\sum_{\bf k}\bar{\Delta}^{2}_{\rm PG}({\bf k})$.
This normal-state pseudogap $\bar{\Delta}_{\rm PG}({\bf k})$ together with the
related energy dispersion $\varepsilon_{0{\bf k}}$ are obtained straightforwardly
from the electron self-energy $\Sigma_{\rm ph}({\bf k},\omega)$. In previous
studies \cite{Feng12,Li25}, the evolution of $\bar{\Delta}_{\rm PG}$ with doping
has been discussed, where the obtained result of the normal-state pseudogap
magnitude at an any given doping is well consistent with the experimentally
measured corresponding normal-state pseudogap energy scale
\cite{Timusk99,Hufner08,Hussey11,Vishik18}, while the correlation of the
normal-state pseudogap magnitude to doping (then the phase diagram) is also in
agreement with the experimental observations
\cite{Timusk99,Hufner08,Hussey11,Vishik18}, i.e., $\bar{\Delta}_{\rm PG}$ is
relatively large at the slight underdoping, and then it decreases with the increase
of doping in the underdoped regime, eventually terminating at the strongly overdoped
region. On the other hand, at a given doping concentration, $\bar{\Delta}_{\rm PG}$
is defined as a crossover with a pseudogap crossover temperature $T^{*}$, where in
corresponding to the doping dependence of $\bar{\Delta}_{\rm PG}$, $T^{*}$ decreases
with the increase of doping, and goes approximately to zero in the strongly
overdoped region \cite{Timusk99,Hufner08,Hussey11,Vishik18}.

\begin{figure}[h!]
\centering
\includegraphics[scale=0.85]{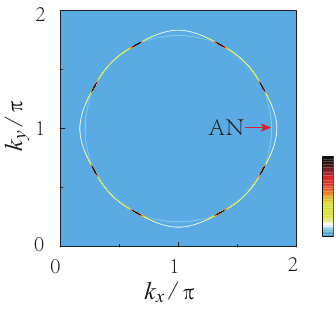}
\caption{(Color online) The intensity map of the electron spectral function in the first
Brillouin zone for binding-energy $\omega=0$ at $\delta=0.09$ with $T=0.002J$, where the
Brillouin zone center has been shifted by [$\pi,\pi$], while AN labelled the antinode.
\label{EFS-maps}}
\end{figure}

$T^{*}$ is actually a crossover temperature below which a novel electronic state
emerges, where the distinguished features are characterized by the presence of the
ordering phenomena \cite{Comin16,Comin14,Gh12,Neto14,Campi15,Hash15}, the dramatic
change in the line-shape of the energy distribution curve
\cite{Hash15,Dessau91,Campuzano99,Lu01,Sato02,Borisenko03,Wei08,Loret17,DMou17}, and
the kink in the quasiparticle dispersion
\cite{Bogdanov00,Kaminski01,Johnson01,Sato03,Lee09-1,He13}, etc. In particular, the
normal-state pseudogap leads to an EFS reconstruction
\cite{Shi08,Sassa11,Horio16,Loret18,Chatterjee06,McElroy06,Chatterjee07,He14,Restrepo23}.
The location of the underlying EFS contour is obtained straightforwardly by the
poles of the full electron propagator (\ref{EGF}) at zero energy, i.e.,
$\varepsilon_{\bf k}+{\rm Re}\Sigma_{\rm ph}({\bf k},0)=\bar{\varepsilon}_{\bf k}=0$,
and then the spectral weight at the EFS contour is governed mainly by the imaginary
part of the electron self-energy ${\rm Im}\Sigma_{\rm ph}({\bf k},\omega)$, where
$\bar{\varepsilon}_{\bf k}=Z_{\rm F}\varepsilon_{\bf k}$ is the renormalized
electron energy dispersion and
$Z^{-1}_{\rm F}=1-{\rm Re}\Sigma_{\rm pho}({\bf k},0)\mid_{{\bf k}=[\pi,0]}$ is the
single-particle coherent weight, with ${\rm Re}\Sigma_{\rm pho}({\bf k},\omega)$
that is the real part of the antisymmetric part of the electron self-energy. In 
the previous works \cite{Li25,Feng15a,Feng16}, the EFS reconstruction has been discussed
in the detail. Since the shape of EFS plays a crucial role in the understanding of
the low-temperature resistivity in the normal-state of cuprate superconductors, we
{\it replot} the intensity map of the electron spectral function
$A({\bf k},\omega)=-{\rm Im}G({\bf k},\omega)/\pi$ in the first Brillouin zone for
binding-energy $\omega=0$ at $\delta=0.09$ with $T=0.002J$ in Fig. \ref{EFS-maps},
where the Brillouin zone center has been shifted by [$\pi,\pi$], while AN labelled
the antinode. 
It thus shows that the normal-state pseudogap generates an EFS
reconstruction, i.e., the spectral weight at around the antinodal region of EFS
becomes partially gapped, leading to that EFS consists, not of a closed contour,
but only of four disconnected Fermi arcs centered at around the nodal region, in
qualitative agreement with the experimental observations
\cite{Shi08,Sassa11,Horio16,Loret18,Chatterjee06,McElroy06,Chatterjee07,He14,Restrepo23}.

\subsection{Spin Pseudogap and Full Effective Spin Propagator}
\label{Effective-spin-propagator}

In this subsection, we study the detailed features of the full effective spin
propagator and the related spin pseudogap in the fermion-spin theory (\ref{CSSFS})
description of the low-energy effective $t$-$J$ model (\ref{tJ-model}) for a
convenience in the following discussion of the low-temperature resistivity due to
the impurity-scattering assisted electron umklapp scattering mediated by the
exchange of the full effective spin propagator. The full effective spin propagator
can be expressed as \cite{Ma24},
\begin{eqnarray}\label{ESP-1}
P({\bf k},{\bf p},{\bf k}',\omega)={1\over N}\sum_{\bf q}
\Lambda_{{\bf p}+{\bf q}+{\bf k}}
\Lambda_{{\bf q}+{\bf k}'}\bar{\Pi}({\bf p},{\bf q},\omega),
\end{eqnarray}
with the full spin bubble,
\begin{eqnarray}\label{spin-bubble-1}
\bar{\Pi}({\bf p},{\bf q},ip_{m})={1\over\beta}\sum_{iq_{m}}D({\bf q},iq_{m})
D({\bf q}+{\bf p},iq_{m}+ip_{m}),~~~~
\end{eqnarray}
where the full spin propagator $D({\bf k},\omega)$ can be expressed
as\cite{Kuang15,Yuan01,Feng98},
\begin{eqnarray}\label{FSGF}
D({\bf k},\omega)={1\over D^{(0)-1}({\bf k},\omega)
-\Sigma^{({\rm s})}_{\rm ph}({\bf k},\omega)},~~~~~
\end{eqnarray}
with the spin self-energy $\Sigma^{({\rm s})}_{\rm ph}({\bf k},\omega)$, which is
derived in terms of the collective charge-carrier mode, and has been given in
Ref. \onlinecite{Kuang15}. Starting from this full spin propagator (\ref{FSGF}), the
dynamical spin response of cuprate superconductors has been investigated
\cite{Kuang15,Yuan01,Feng98}, and the obtained results are qualitatively consistent
with the experimental observations
\cite{Birgeneau89,Fong95,Yamada98,Arai99,Bourges00,He01,Tranquada04,Bourges05}.

As in the case of the normal-state pseudogap discussed in subsection
\ref{Phase-diagram}, the spin pseudogap $\bar{\Delta}^{({\rm s})}_{\rm pg}({\bf k})$
in the fermion-spin theory (\ref{CSSFS}) description of the $t$-$J$ model
(\ref{tJ-model}) is closely related to the spin self-energy
$\Sigma^{({\rm s})}_{\rm ph}({\bf k},\omega)$ as\cite{Ma24},
\begin{equation}\label{SSFN-spin-gap}
\Sigma^{({\rm s})}_{\rm ph}({\bf k},\omega)
\approx {B_{\bf k}[\bar{\Delta}^{({\rm s})}_{\rm pg}({\bf k})]^{2}\over \omega^{2}
-\omega^{2}_{0{\bf k}}},
\end{equation}
where $\bar{\Delta}^{({\rm s})}_{\rm pg}({\bf k})$ anisotropically suppresses the
spin excitation density of states, which together with the related spin excitation
energy dispersion $\omega_{0{\bf k}}$ can be obtained directly from the spin
self-energy $\Sigma^{({\rm s})}_{\rm ph}({\bf k},\omega)$, and have been given in
Ref. \onlinecite{Ma24}. Substituting the above spin self-energy (\ref{SSFN-spin-gap}) 
into Eq. (\ref{FSGF}), the full spin propagator (\ref{FSGF}) can be obtained as,
\begin{eqnarray}\label{FSGF-spin-gap}
D({\bf k},\omega)={\bar{B}_{1{\bf k}}\over\omega^{2}-\bar{\omega}^{2}_{1{\bf k}}}
+{\bar{B}_{2{\bf k}}\over\omega^{2}-\bar{\omega}^{2}_{2{\bf k}}}
=\sum_{\alpha=1,2}{\bar{B}_{\alpha{\bf k}}\over\omega^{2}
-\bar{\omega}^{2}_{\alpha{\bf k}}},~~~~~~
\end{eqnarray}
with the renormalized spin excitation energy dispersions,
\begin{subequations}\label{FSEED}
\begin{eqnarray}
\bar{\omega}^{2}_{1{\bf k}}&=&{1\over 2}\left [\omega^{2}_{\bf k}
+\omega^{2}_{0{\bf k}}+\sqrt{(\omega^{2}_{\bf k}-\omega^{2}_{0{\bf k}})^{2}
+4B^{2}_{\bf k}[\bar{\Delta}^{({\rm s})}_{\rm pg}({\bf k})]^{2}}\right ],\nonumber\\
~~~~~~ \\
\bar{\omega}^{2}_{2{\bf k}}&=&{1\over 2}\left [\omega^{2}_{\bf k}
+\omega^{2}_{0{\bf k}}-\sqrt{(\omega^{2}_{\bf k}-\omega^{2}_{0{\bf k}})^{2}
+4B^{2}_{\bf k}[\bar{\Delta}^{({\rm s})}_{\rm pg}({\bf k})]^{2}}\right ],\nonumber\\
~~~~~~
\end{eqnarray}
\end{subequations}
and the corresponding spectral weights $\bar{B}_{1{\bf k}}$ and
$\bar{B}_{2{\bf k}}$, respectively, which have been derived in Ref. \onlinecite{Ma24}.

\begin{figure}[h!]
\centering
\includegraphics[scale=0.65]{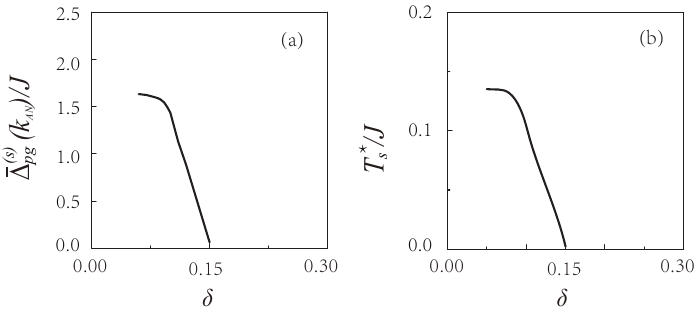}
\caption{(a) The antinodal spin pseudogap
$\bar{\Delta}^{({\rm s})}_{\rm pg}({\bf k}_{\rm AN})$ in temperature $T=0.002J$ and
(b) the corresponding antinodal spin pseudogap crossover temperature $T^{*}_{\rm s}$
as a function of doping, where ${\bf k}_{\rm AN}$ is the wave vector at the antinode.
\label{spin-pseudogap-doping}}
\end{figure}

The crucial scattering process responsible for the low-temperature resistivity in
the overdoped strange-metal phase is concentrated at around the antinodal region,
however, the antinodal spin pseudogap opens at low temperatures in the underdoped
pseudogap phase, which partially suppresses the spin excitation density of states
at around the antinodal region \cite{Bucher93,Ito93,Nakano94,Ando01}, and thus
would naturally account for a deviation from the low-temperature behaviour of the
resistivity in the overdoped strange-metal phase.
In Fig. \ref{spin-pseudogap-doping}a, we {\it replot} the antinodal spin pseudogap
$\bar{\Delta}^{({\rm s})}_{\rm pg}({\bf k}_{\rm AN})$ as a function of doping with
temperature $T=0.002J$, where ${\bf k}_{\rm AN}$ is the wave vector at the antinode
of EFS. The obtained result\cite{Ma24} in Fig. \ref{spin-pseudogap-doping}a therefore
shows that $\bar{\Delta}^{({\rm s})}_{\rm pg}({\bf k}_{\rm AN})$ is not sensitive to
doping in the slightly underdoped region, and then it decreases rapidly as doping is
increased in the heavily underdoped region. More specially,
$\bar{\Delta}^{({\rm s})}_{\rm pg}({\bf k}_{\rm AN})$ vanishes abruptly at around the
optimal doping, indicating that the main properties of the antinodal spin excitation
in the overdoped regime can be well described by the MF spin propagator (\ref{SGF-1}).
This $\bar{\Delta}^{({\rm s})}_{\rm pg}({\bf k}_{\rm AN})$ in the underdoped regime
depresses the spin excitation density of states at around the antinodal region.
Moreover, for a given doping, $\bar{\Delta}^{({\rm s})}_{\rm pg}({\bf k}_{\rm AN})$
vanishes when the temperature reaches the antinodal spin pseudogap crossover
temperature $T^{*}_{\rm s}$. 
In Fig. \ref{spin-pseudogap-doping}b, we {\it replot}
$T^{*}_{\rm s}$ as a function of doping, where $T^{*}_{\rm s}$ presents a similar
behavior of $\bar{\Delta}^{({\rm s})}_{\rm pg}({\bf k}_{\rm AN})$, i.e.,
$T^{*}_{\rm s}$ is relatively high in the slightly underdoped region, however, it
diminishes quickly with the increase of doping in the heavily underdoped region, and
terminates eventually at around the optimal doping.

From the above full spin propagator (\ref{FSGF-spin-gap}), the full spin bubble
(\ref{spin-bubble-1}) can be evaluated as,
\begin{equation}\label{spin-bubble}
\bar{\Pi}({\bf p},{\bf q},\omega)=-\sum_{\substack{\alpha=1,2\\ \alpha'=1,2}}
{\bar{W}^{(1)}_{\alpha\alpha'{\bf p}{\bf q}}\over\omega^{2}
-[\bar{\omega}^{(1)}_{\alpha\alpha'{\bf p}{\bf q}}]^{2}}
+{\bar{W}^{(2)}_{\alpha\alpha'{\bf p}{\bf q}}\over\omega^{2}
-[\bar{\omega}^{(2)}_{\alpha\alpha'{\bf p}{\bf q}}]^{2}},~~~~~~
\end{equation}
with the full effective spin excitation energy dispersions,
\begin{subequations}\label{effecive-SEED}
\begin{eqnarray}
\bar{\omega}^{(1)}_{\alpha\alpha'{\bf p}{\bf q}}
&=&\bar{\omega}_{\alpha{\bf q}+{\bf p}}+\bar{\omega}_{\alpha'{\bf q}}, \\
\bar{\omega}^{(2)}_{\alpha\alpha'{\bf p}{\bf q}}
&=&\bar{\omega}_{\alpha{\bf q}+{\bf p}}-\bar{\omega}_{\alpha'{\bf q}},
\end{eqnarray}
\end{subequations}
and the corresponding weight functions $\bar{W}^{(1)}_{\alpha\alpha'{\bf p}{\bf q}}$
and $\bar{W}^{(2)}_{\alpha\alpha'{\bf p}{\bf q}}$, respectively, which have been
derived in Ref. \onlinecite{Ma24}. Substituting the above full spin bubble
(\ref{spin-bubble}) into Eq. (\ref{ESP-1}), the full effective spin propagator
(\ref{ESP-1}) can be obtained as,
\begin{eqnarray}\label{reduced-propagator}
P({\bf k},{\bf p},{\bf k}',\omega) &=& -{1\over N}\sum\limits_{\alpha\alpha'{\bf q}}
\left [ {\varpi^{(1)}_{\alpha\alpha'}({\bf k},{\bf p},{\bf k}',{\bf q})\over
\omega^{2}-[\bar{\omega}^{(1)}_{\alpha\alpha'{\bf p}{\bf q}}]^{2}}\right .\nonumber\\
&-&\left . {\varpi^{(2)}_{\alpha\alpha'}({\bf k},{\bf p},{\bf k}',{\bf q})\over
\omega^{2}-[\bar{\omega}^{(2)}_{\alpha\alpha'{\bf p}{\bf q}}]^{2}} \right ],~~~~~
\end{eqnarray}
where the corresponding spectral weights
$\varpi^{(1)}_{\alpha\alpha'}({\bf k},{\bf p},{\bf k}',{\bf q})=
\Lambda_{{\bf k}+{\bf p}+{\bf q}}\Lambda_{{\bf q}+{\bf k}'}
\bar{W}^{(1)}_{\alpha\alpha'{\bf p}{\bf q}}$ and
$\varpi^{(2)}_{\alpha\alpha'}({\bf k},{\bf p},{\bf k}',{\bf q})=
\Lambda_{{\bf k}+{\bf p}+{\bf q}}\Lambda_{{\bf q}+{\bf k}'}
\bar{W}^{(2)}_{\alpha\alpha'{\bf p}{\bf q}}$,
respectively.

We now turn to explore the exotic properties of the full spin propagator
(\ref{reduced-propagator}). The full spin excitation spectral function
$A_{\rm spin}({\bf k},\omega)= -{\rm Im}D({\bf k},\omega)/\pi$ can be obtained
straightforwardly from the full spin propagator (\ref{FSGF-spin-gap}) as,
\begin{eqnarray}
A_{\rm spin}({\bf k},\omega) = A^{(1)}_{\rm spin}({\bf k},\omega)
+A^{(2)}_{\rm spin}({\bf k},\omega),
\end{eqnarray}
with the corresponding components,
\begin{subequations}\label{full-spectral-function}
\begin{eqnarray}
A^{(1)}_{\rm spin}({\bf k},\omega)&=&{\bar{B}_{1{\bf k}}\over
\bar{\omega}_{1{\bf k}}}
[\delta(\omega-\bar{\omega}_{1{\bf k}})-\delta(\omega+\bar{\omega}_{1{\bf k}})],
~~~~\label{full-spectral-function-1}\\
A^{(2)}_{\rm spin}({\bf k},\omega)&=& {\bar{B}_{2{\bf k}}\over
\bar{\omega}_{2{\bf k}}}
[\delta(\omega-\bar{\omega}_{2{\bf k}})-\delta(\omega+\bar{\omega}_{2{\bf k}})]
\label{full-spectral-function-2}.
\end{eqnarray}
\end{subequations}
However, after a careful calculation and analysis \cite{Ma24}, it has been found that
the spectral weight of the full spin excitation spectrum in the component
$A^{(1)}_{\rm spin}({\bf k},\omega)$ is several orders of magnitude larger than the
corresponding spectral weight in the component $A^{(2)}_{\rm spin}({\bf k},\omega)$,
which shows that the constrained electrons in the two-dimensional Fermi sea are mainly
scattered by the spin excitations from the component
$A^{(1)}_{\rm spin}({\bf k},\omega)$. On the other hand, it has been also found that
the highest density of states of the spin excitations with the renormalized spin
excitation energy dispersion $\bar{\omega}_{1{\bf k}}$ is located at around the AF
wave vector ${\bf k}_{\rm A}=[\pm\pi,\pm\pi]$. With the help of these special
properties of the spin excitations, the full effective spin propagator
$P({\bf k},{\bf p},{\bf k}',\omega)$ in Eq. (\ref{reduced-propagator}) now can be
reduced approximately as,
\begin{eqnarray}\label{reduced-propagator-1}
P({\bf k},{\bf p},{\bf k}',\omega)&\approx& -{1\over N}\sum\limits_{\bf q}\left [
{\varpi^{(1)}_{11}({\bf k},{\bf p},{\bf k}',{\bf q})\over\omega^{2}
-[\omega^{(1)}_{11{\bf p}{\bf q}}]^{2}}\right. \nonumber\\
&-&\left . {\varpi^{(2)}_{11}({\bf k},{\bf p},{\bf k}',{\bf q})\over\omega^{2}
-[\omega^{(2)}_{11{\bf p}{\bf q}}]^{2}} \right ].~~~~~
\end{eqnarray}
Following previous discussion\cite{Ma23,Ma24}, the full effective spin excitation
energy dispersions $\bar{\omega}^{(1)}_{11{\bf p}{\bf q}}$ and
$\bar{\omega}^{(2)}_{11{\bf p}{\bf q}}$ in Eq. (\ref{effecive-SEED}) can be expressed
approximately in terms of the taylor expansion as,
\begin{subequations}\label{effective-spin-excitation}
\begin{eqnarray}
\bar{\omega}^{(1)}_{11{\bf p}{\bf q}}&=&\bar{\omega}_{1{\bf q}+{\bf p}}
+\bar{\omega}_{1{\bf q}}\approx b_{\bf q}p^{2}+2\bar{\omega}_{1{\bf q}},\\
\bar{\omega}^{(2)}_{11{\bf p}{\bf q}}&=&\bar{\omega}_{1{\bf q}+{\bf p}}
-\bar{\omega}_{1{\bf q}}\approx b_{\bf q}p^{2},
\end{eqnarray}
\end{subequations}
with $b_{\bf q}=d^{2}\bar{\omega}_{1{\bf q}}/(d^{2}{\bf q})$. It thus shows that the
full effective spin excitation energy in Eq. (\ref{reduced-propagator-1}) scales with
$p^{2}$.

\subsection{Boltzmann Transport Equation}\label{Boltzmann-theory}

For the discussion of the nature of the low-temperature resistivity of cuprate
superconductors in the normal-state, we need to determine the momentum distribution
relaxation. A popular method for the calculation of the momentum distribution
relaxation is to solve the Boltzmann transport equation with the input of the
scattering processes\cite{Abrikosov88,Mahan81}, since the Boltzmann transport
equation is effective in the presence of well-defined quasiparticles or in dealing
with electron interactions mediated by different boson modes in the Eliashberg
approach. This is based on the groundbreaking work of Prange and Kadanoff
\cite{Prange64}, who demonstrated that in the electron-phonon system, a set of
transport equations can be derived under the Migdal approximation, where the
phonon mediated electron interaction leads to the electron self-energy and vertex
correction. Especially, even in the absence of clearly defined quasiparticles, this
set of coupled transport equations for electron and phonon distribution functions
is correct, where one of the forms of the
transport equation \cite{Prange64},
\begin{eqnarray}\label{Boltzmann-equation-2}
e{\bf E}\cdot\nabla_{\bf k}f({\bf k})=-I_{\rm e-e}-I_{\rm i-e},
\end{eqnarray}
with the input of the scattering process is identical to the electrical Boltzmann
transport equation proposed very early by Landau for the case in which the
quasiparticle is well-defined \cite{Abrikosov88,Mahan81}, where $e$ is the charge of
an electron, ${\bf E}$ is an external electric field ${\bf E}$, $f({\bf k},t)$ is
the electron distribution function in a homogeneous system, $I_{\rm e-e}$ is the
electron-electron collision term, and $I_{\rm i-e}$ is the electron-impurity
collision term. More importantly, it has been shown clearly that the Boltzmann
transport equation derived by Prange and Kadanoff \cite{Prange64} is not specific to
an electron interaction mediated by phonons in the electron-phonon system, and is
also effective for the system with the electron interaction mediated by other boson
modes \cite{Lee21}.

In this paper, we start from the microscopic electronic structure of cuprate
superconductors discussed in subsection \ref{Phase-diagram} to study the
low-temperature resistivity in the normal-state by solving the Boltzmann transport
equation (\ref{Boltzmann-equation-2}). To solve this Boltzmann transport equation
(\ref{Boltzmann-equation-2}), the linear perturbation from the equilibrium in terms
of the fermion distribution function $n_{\rm F}(\omega)$ and the local shift of the
chemical potential $\tilde{\Phi}({\bf k})$ has been introduced \cite{Prange64,Lee21}
as, $f({\bf k})= n_{\rm F}({\bar{\varepsilon}_{\bf k}})-
[dn_{\rm F}({\bar{\varepsilon}_{\bf k}})/d{\bar{\varepsilon}_{\bf k}}]
\tilde{\Phi}({\bf k})$, where $\tilde{\Phi}({\bf k})$ satisfies the antisymmetric
relation $\tilde{\Phi}(-{\bf k})=-\tilde{\Phi}({\bf k})$. With the help of the above
treatment, the Boltzmann transport equation (\ref{Boltzmann-equation-2}) can
linearized as,
\begin{eqnarray}\label{Boltzmann-equation-3}
e{\bf v}_{\bf k}\cdot{\bf E}{dn_{\rm F}({\bar{\varepsilon}_{\bf k}})\over
d{\bar{\varepsilon}_{\bf k}}}=-I_{\rm e-e}-I_{\rm i-e},
\end{eqnarray}
where the momentum dependence of the electron velocity
${\bf v}_{\bf k}=\nabla_{\bf k}{\bar{\varepsilon}_{\bf k}}$.

The electron-impurity collision term $I_{\rm i-e}$ in the Boltzmann transport
equation (\ref{Boltzmann-equation-3}) is proportional to the local shift of the
chemical potential $\tilde{\Phi}({\bf k})$, and it is straightforward to find the
electron-impurity collision term $I_{\rm i-e}$ as \cite{Lee21},
\begin{eqnarray}\label{impurity-collision}
I_{\rm i-e}=-\gamma_{0}\tilde{\Phi}({\bf k}){dn_{\rm F}({\bar{\varepsilon}_{\bf k}})
\over d\bar{\varepsilon}_{\bf k}},
\end{eqnarray}
with the impurity scattering rate $\gamma_{0}$. This impurity scattering is required
to restrict the modification of the distribution function to at around the antinodal
region [see Appendix \ref{NSBE}]. On the other hand, the electron-electron collision
term $I_{\rm e-e}$ in the Boltzmann transport equation (\ref{Boltzmann-equation-3})
is closely related to the mechanism of the momentum relaxation
\cite{Abrikosov88,Mahan81}. In this paper, we adopt the impurity-scattering assisted
electron umklapp scattering as the mechanism of the momentum relaxation
\cite{Lee21,Rice17,Honerkamp01,Hartnoll12,Hussey03}. The reasons can be summarized as:
(i) as shown in Fig. \ref{EFS-maps}, the spectral weight on EFS at around the antinodal
region is suppressed partially by the normal-state pseudogap to form the Fermi arcs,
which therefore shows that the interaction (then the scattering) between electrons
at around the antinodal region is particularly strong, and then the important
impurity-scattering assisted electron umklapp scattering is concentrated at around
the antinodal region; (ii) the minimal electron umklapp vector $\Delta_{p}$ connecting
neighboring EFS is small at around the antinodal region \cite{Lee21}. In particular,
in the present impurity-scattering assisted electron umklapp scattering from a spin
excitation, the temperature scale proportional to $\Delta^{2}_{p}$ presents a similar
behavior of the antinodal spin pseudogap crossover temperature in the underdoped
pseudogap phase \cite{Ma24}, however, it can be very low in the overdoped
strange-metal phase. We will turn to further discuss this issue towards in
Section \ref{electron-resistivity}.

In our previous discussions \cite{Ma23,Ma24}, the electron-electron collision
$I_{\rm e-e}$ in the Boltzmann transport equation (\ref{Boltzmann-equation-3}) due to
the impurity-scattering assisted electron umklapp scattering from a spin excitation
has been derived, and can be expressed explicitly as,
\begin{widetext}
\begin{eqnarray}\label{electron-collision-1}
I_{\rm e-e}&=&{1\over N^{2}}\sum_{{\bf k}',{\bf p}} {2\over T}
|{\bar P}({\bf k},{\bf p},{\bf k}',\bar{\varepsilon}_{\bf k}
-\bar{\varepsilon}_{{\bf k}+{\bf p}+{\bf G}})|^{2}
\{\tilde{\Phi}({\bf k})+\tilde{\Phi}({\bf k'})
-\tilde{\Phi}({\bf k}+{\bf p}+{\bf G})-\tilde{\Phi}({\bf k}'-{\bf p})\}\nonumber\\
&\times& n_{\rm F}(\bar{\varepsilon}_{\bf k})
n_{\rm F}(\bar{\varepsilon}_{{\bf k}'})
[1-n_{\rm F}(\bar{\varepsilon}_{{\bf k}+{\bf p}+{\bf G}})]
[1-n_{\rm F}(\bar{\varepsilon}_{{\bf k}'-{\bf p}})]
\delta(\bar{\varepsilon}_{\bf k}+\bar{\varepsilon}_{\bf k'}
-\bar{\varepsilon}_{{\bf k}+{\bf p}+{\bf G}}-\bar{\varepsilon}_{{\bf k}'-{\bf p}}),
\end{eqnarray}
\end{widetext}
where ${\bf G}$ represents a set of reciprocal lattice vectors, and following the
common practice\cite{Ma23}, the scattering probability for two electrons has been
normalized as \linebreak  ${\bar P}({\bf k},{\bf p},{\bf k}',\bar{\varepsilon}_{\bf k}
-\bar{\varepsilon}_{{\bf k}+{\bf p}+{\bf G}})=
P({\bf k},{\bf p},{\bf k}',\bar{\varepsilon}_{\bf k}
-\bar{\varepsilon}_{{\bf k}+{\bf p}+{\bf G}})/W_{\rm sp}$ with the normalization
factor \linebreak  $W^{2}_{\rm sp}=(1/N^{2})\sum_{{\bf k},{\bf p}}
\int |{\rm Im}\bar{P}({\bf k},{\bf p}-{\bf k},\omega)|^{2}d\omega$. It is worth
noting that (i) the above electron umklapp scattering (\ref{electron-collision-1})
is described as a scattering between electrons by the exchange of the full effective
spin propagator $P({\bf k},{\bf p},{\bf k}',\omega)$ in Eq. (\ref{ESP-1}), rather
than the scattering between electrons via the emission and absorption of the spin
excitation \cite{Lee21}; (ii) the above electron umklapp scattering process
(\ref{electron-collision-1}) is a rather special scattering process, since small
momentum scattering is focused on only, and in this case, this mechanism of the
momentum relaxation requires both the electron umklapp scattering and impurity
scattering\cite{Lee21} [see Appendix \ref{NSBE}], although the low-temperature
resistivity slope is independent of the impurity scattering.

In an interacting electron system, all the low-temperature transport processes
involve only the electronic states near EFS \cite{Abrikosov88,Mahan81}. In this
case, a given patch at EFS can be described by the Fermi angle $\theta$ with
$\theta\in [0,2\pi]$. In the present case of the impurity-scattering
assisted umklapp scattering between electrons by the exchange of the full effective
spin propagator, an electron on EFS parameterized by the Fermi angle $\theta$ is
scattered to a point parameterized by the Fermi angle $\theta'$ on the umklapp EFS
via the spin excitation carrying momentum \cite{Ma23,Ma24}, and then the momentum
integration in the electron-electron collision (\ref{electron-collision-1}) along
the perpendicular direction can be replaced by the integration over
$\bar{\varepsilon}_{\bf k}$ \cite{Prange64,Lee21}.  With the help of the above
treatment, the electron-electron collision $I_{\rm e-e}$ in
Eq. (\ref{electron-collision-1}) has been obtained \cite{Ma23}, and then the Boltzmann
transport equation (\ref{Boltzmann-equation-3}) can be further expressed as,
\begin{eqnarray}\label{electron-collision}
e{\bf v}_{\rm F}(\theta)\cdot {\bf E}&=&-[\gamma(\theta)+\gamma_{0}]\Phi(\theta)
\nonumber\\
&+& 2\int {d\theta'\over {2\pi}}\zeta(\theta')F(\theta,\theta')\Phi(\theta'),~~~~~
\end{eqnarray}
with $\Phi(\theta)=\tilde{\Phi}[{\rm k}(\theta)]$, where the antisymmetric relation
$\tilde{\Phi}(-{\bf k})=-\tilde{\Phi}({\bf k})$ for $\tilde{\Phi}({\bf k})$ has been
replaced as $\Phi(\theta)=-\Phi(\theta+\pi)$ for $\Phi(\theta)$, the Fermi velocity
${\bf v}_{\rm F}(\theta)$ at the Fermi angle $\theta$, the density of states factor
$\zeta(\theta')={\rm k}^{2}_{\rm F}/[4\pi^{2}{\rm v}^{3}_{\rm F}]$ at angle
$\theta'$, the Fermi wave vector ${\rm k}_{\rm F}$, the Fermi velocity
${\rm v}_{\rm F}$, and the angular (momentum) dependence of the electron umklapp
scattering rate,
\begin{eqnarray}\label{scattering-rate}
\gamma(\theta)=2\int {d \theta' \over {2\pi}} \zeta(\theta')F(\theta,\theta'),
\end{eqnarray}
where the impurity-scattering assisted kernel function $F(\theta,\theta')$ connecting
the point $\theta$ on the circular EFS with the point $\theta'$ on the umklapp EFS
via the amplitude of the momentum transfer ${\rm p}(\theta,\theta')$ can be expressed
as,
\begin{eqnarray}\label{kernel-function}
F(\theta,\theta')&=& {1\over T}\int {d\omega\over 2\pi}{\omega^{2}\over
{\rm p}(\theta,\theta')}
{|\bar{P}[{\rm k}(\theta),{\rm p}(\theta,\theta'),\omega]|}^{2}\nonumber\\
&\times& n_{\rm B}(\omega)[1+n_{\rm B}(\omega)],~~~~~~
\end{eqnarray}
where $n_{\rm B}(\omega)$ is the boson distribution function, while the full
effective spin propagator $\bar{P}({\bf k},{\bf p},{\bf k}',\omega)$
in Eq. (\ref{electron-collision-1}) can be further expressed in terms of the
Fermi angles $\theta$ and $\theta'$ as\cite{Ma24},
\begin{eqnarray}\label{reduced-propagator-5}
\bar{P}[{\rm k}(\theta),{\rm p}(\theta,\theta'),\omega]&=&-{1\over W_{\rm sp}}
{1\over N}\sum\limits_{\alpha\alpha'{\bf q}}\left [
{\varpi^{(1)}_{\alpha\alpha'}(\theta,\theta',{\bf q})\over\omega^{2}
-[\bar{\omega}^{(1)}_{\alpha\alpha'\theta,\theta'}({\bf q})]^{2}}\right .
\nonumber\\
&-&\left . {\varpi^{(2)}_{\alpha\alpha'}(\theta,\theta',{\bf q})\over\omega^{2}
-[\bar{\omega}^{(2)}_{\alpha\alpha'\theta,\theta'}({\bf q})]^{2}} \right ],
\end{eqnarray}
with $\varpi^{(1)}_{\alpha\alpha'}(\theta,\theta',{\bf q})=
\varpi^{(1)}_{\alpha\alpha'}
[{\rm k}(\theta),{\rm p}(\theta,\theta'),{\bf k}'_{\rm F},{\bf q}]$,
$\varpi^{(2)}_{\alpha\alpha'}(\theta,\theta',{\bf q})=
\varpi^{(2)}_{\alpha\alpha'}
[{\rm k}(\theta),{\rm p}(\theta,\theta'),{\bf k}'_{\rm F},{\bf q}]$,
$\bar{\omega}^{(1)}_{\alpha\alpha'\theta,\theta'}({\bf q})
=\bar{\omega}^{(1)}_{\alpha\alpha'{\rm p}(\theta,\theta'){\bf q}}$, and
$\bar{\omega}^{(2)}_{\alpha\alpha'\theta,\theta'}({\bf q})
=\bar{\omega}^{(2)}_{\alpha\alpha'{\rm p}(\theta,\theta'){\bf q}}$.

\section{Low-Temperature Resistivity}\label{electron-resistivity}

From the Boltzmann transport equation (\ref{electron-collision}), the electron
current density can be obtained in terms of the local shift of the chemical
potential as \cite{Ma23,Ma24},
\begin{eqnarray}\label{current-density}
{\bf J} &=& -en_{0}{1\over N}\sum_{\bf k}{\bf v}_{\bf k}
{dn_{\rm F}({\bar{\varepsilon}_{\bf k}})\over d\bar{\varepsilon}_{\bf k}}
\tilde{\Phi}({\bf k})\nonumber\\
&=& -en_{0}{{\rm k}_{\rm F}\over {\rm v}_{\rm F}}\int
{d\theta\over (2\pi)^{2}}{\bf v}_{\rm F}(\theta)\Phi(\theta),~~~~~
\end{eqnarray}
with the momentum relaxation that is generated by the action of the electric field
on the mobile electrons at EFS with the density $n_{0}$.

The local shift of the chemical potential $\Phi(\theta)$ in the Boltzmann transport
equation (\ref{electron-collision}) can be derived numerically [see Appendix
\ref{NSBE}] or evaluated directly in the relaxation-time approximation. In particular,
starting from the impurity-scattering assisted electron umklapp scattering mediated
by a critical boson mode, the accurate solution and the approximated result for the
local shift of the chemical potential are respectively obtained from numerically
solving the corresponding inverse matrix and in the relaxation time approximation
\cite{Lee21}, respectively. Comparing the results of the resistivity obtained in the
relaxation time approximation with the corresponding results obtained from the
accurate solution for the local shift of the chemical potential, it is thus shown
that the relaxation time approximation works very well at low temperatures. In this
case, as a qualitative discussion in this paper, we discuss the
low-temperature resistivity due to the impurity-scattering assisted electron umklapp
scattering from a spin excitation in the relaxation time approximation only, while
the formalism for the numerical solution of the local shift of the chemical potential
is presented in Appendix \ref{NSBE} for the emphasis of the mechanism of the momentum
relaxation from any a boson mode requires both the impurity scattering and the
electron umklapp scattering.

In the relaxation-time approximation, the local shift of the chemical potential
$\Phi(\theta)$ in the electron current density equation (\ref{current-density}) can
be evaluated straightforwardly as,
\begin{eqnarray}\label{solution-Phi-RTA}
\Phi(\theta)=-{e{\rm v}_{\rm F}{\rm cos}(\theta)\over 2\gamma(\theta)
+\gamma_{0}}E_{\hat{x}},
\end{eqnarray}
where the electric field ${\bf E}$ has been chosen along the $\hat{x}$-axis, and
then the dc conductivity now can be obtained as
\cite{Lee21,Ma23},
\begin{eqnarray}\label{dc-conductivity}
\sigma_{\rm dc}(T)=e^{2}n_{0}{\rm k}_{\rm F}{\rm v}_{\rm F}\int {d\theta\over
(2\pi)^{2}}{\rm cos}^{2}(\theta){1\over 2\gamma(\theta)+\gamma_{0}}. ~~~~
\end{eqnarray}
The above obtained dc conductivity allows us to determine the resistivity in a
straightforward way as,
\begin{eqnarray}\label{dc-resistivity}
\rho(T)={1\over \sigma_{\rm dc}(T)}.
\end{eqnarray}
\begin{figure}[h!]
\centering
\includegraphics[scale=0.85]{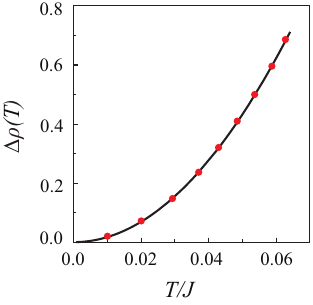}
\caption{(Color online) Low-temperature resistivity as a function of temperature
with the impurity-scattering rate $\gamma_{0}=0.00001$ at the underdoping
$\delta=0.09$, where the red-dots are the numerically fitted results with the fit
form $\Delta\rho(T)=A_{2}T^{2}$ and $A_{2}=173.08$.
\label{resistivity-temperature-underdoping}}
\end{figure}
In particular, as we have shown in previous works \cite{Ma23,Ma24}, when the
temperature $T\rightarrow 0$, the angular (momentum) dependence of the electron
umklapp scattering rate $\gamma(\theta,T)\rightarrow 0$, and then the
impurity-scattering generated resistivity $\rho_{0}$ can be derived directly as
$\rho_{0}=\gamma_{0}/C$, with the temperature independence of the constant,
\begin{eqnarray}\label{impurity-conductivity}
C=e^{2}n_{0}{\rm k}_{\rm F}{\rm v}_{\rm F}\int {d\theta\over (2\pi)^{2}}
{\rm cos}^{2}(\theta)={e^{2}n_{0}{\rm k}_{\rm F}{\rm v}_{\rm F}\over 4\pi}. ~~~~
\end{eqnarray}

\begin{figure}[h!]
\centering
\includegraphics[scale=0.85]{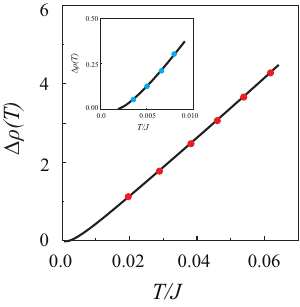}
\caption{(Color online) Low-temperature resistivity as a function of temperature
with the impurity-scattering rate $\gamma_{0}=0.00001$ at the overdoping
$\delta=0.23$, where the red-dots are the numerically fitted results with the fit
form $\Delta\rho(T)=A_{1}T$ and $A_{1}=78.04$. The inset shows the detail of the
temperature dependence of the resistivity in the far-lower-temperature region,
where the blue-dots are the numerically fitted results with the fit form
$\Delta\rho(T)=A_{2}T^{2}$ and $A_{2}=5198.62$.
\label{resistivity-temperature-overdoping}}
\end{figure}

We are now ready to discuss the low-temperature resistivity of cuprate
superconductors in the normal-state. In
Fig. \ref{resistivity-temperature-underdoping}, we plot the low-temperature
resistivity $\Delta\rho(T)=[\rho(T)-\rho_{0}]/\rho_{0}$ as a function of
temperature with the impurity-scattering rate $\gamma_{0}=0.00001$ at the
underdoping $\delta=0.09$, where the red-dots are numerically fitted results with
the fit form $\Delta\rho(T)=A_{2}T^{2}$ and $A_{2}=173.08$. Apparently, in the
low-temperature region of the underdoped pseudogap phase, the low-temperature
resistivity is predominantly T-quadratic, indicating that the characteristic
feature of the T-quadratic behaviour of the low-temperature resistivity is the
same in the theory and experiments
\cite{Bucher93,Ito93,Nakano94,Ando01,Cooper09,Mirzaei13,Barisic13,Pelc20}. Moreover,
we \cite{Ma24} have also shown that the magnitude of the low-temperature T-quadratic
resistivity $\rho(T)$ at a given doping and a given temperature is qualitatively
consistent with the corresponding experimental results in the underdoped pseudogap
phase. On the other hand, the obtained result
of the low-temperature resistivity $\Delta\rho(T)=[\rho(T)-\rho_{0}]/\rho_{0}$ as
a function of temperature with the impurity-scattering rate $\gamma_{0}=0.00001$
at the overdoping $\delta=0.23$ is plotted in
Fig. \ref{resistivity-temperature-overdoping}, where the red-dots are numerically
fitted results with the fit form $\Delta\rho(T)=A_{1}T$ and $A_{1}=78.04$, while
the inset shows the detail of the temperature dependence of the resistivity in
the far-lower-temperature region, where the blue-dots are the numerically fitted
results with the fit form $\Delta\rho(T)=A_{2}T^{2}$ and $A_{2}=5198.62$.
In Fig. \ref{resistivity-temperature-overdoping}, there are two distinguished
regions: (i) in the low-temperature region of the overdoped strange-metal phase,
the low-temperature resistivity presents a T-linear behaviour down to the low
temperature of $T\sim 0.01J$, and (ii) in the far-lower-temperature region
$T< 0.01J$ of the overdoped strange-metal phase, the far-lower-temperature
resistivity decreases quadratically with the decrease of temperature. 
Moreover, we \cite{Ma23,Ma24} have also performed the calculation for the resistivity
$\rho(T)$ at different doping concentrations, and the obtained results show that
the low-temperature T-quadratic resistivity in the underdoped pseudogap phase
persists all the way up to the strongly underdoped region, while the
low-temperature T-linear resistivity in the overdoped strange-metal phase extends
up to the edge of the SC dome, where both the strengths of the low-temperature
T-quadratic resistivity and T-linear resistivity (then the T-quadratic resistivity
and T-linear resistivity coefficients) decrease as doping is increased. 
All these obtained results are well consistent with the corresponding 
experimental observations
\cite{Bucher93,Ito93,Nakano94,Ando01,Cooper09,Mirzaei13,Barisic13,Pelc20,Legros19,Ayres21,Grisso21,Gurvitch87}.
On the other hand, both the slopes of the low-temperature T-quadratic resistivity
and T-linear resistivity themselves are independence of the impurity-scattering
rate $\gamma_{0}$ for $\gamma_{0}\ll 1$, and are therefore intrinsic \cite{Lee21}.
To confirm this point more clearly, we have made a series of calculations for
$\Delta\rho(T)$ at different $\gamma_{0}$, and found that for the case of
$\gamma_{0}< 0.001$, the obtained results of $\Delta\rho(T)$ at different
$\gamma_{0}$ are qualitatively consistent each others.

The above mechanism of the momentum relaxation requires both the impurity scattering
and the electron umklapp scattering, where the impurity scattering is needed to
restrict the modification of the distribution function to the antinodal region
[see Appendix \ref{NSBE}], while the impurity-scattering assisted electron umklapp
scattering from a spin excitation associated with the antinodes leads to the
low-temperature T-linear resistivity in the overdoped strange-metal phase,
however, when this impurity-scattering assisted electron umklapp scattering moves to
the underdoped pseudogap phase, the opening of the antinodal spin pseudogap partially
suppresses the strength of the antinodal electron umklapp scattering, leading to the
low-temperature T-quadratic resistivity. 
\begin{figure}[h!]
\centering
\includegraphics[scale=0.85]{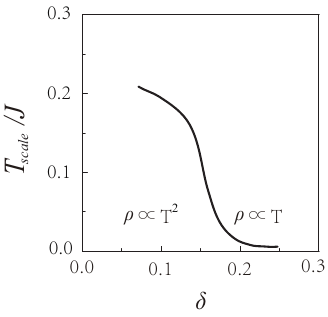}
\caption{Temperature scale $T_{\rm scale}$ as a function of doping.
\label{T-scale-doping}}
\end{figure}
This picture of the unusual crossover from
the low-temperature T-linear resistivity in the overdoped strange-metal phase to the
low-temperature T-quadratic resistivity in the underdoped pseudogap phase can be
also understood from the special properties of the impurity-scattering assisted
kernel function $F(\theta,\theta')$ in Eq. (\ref{kernel-function}) [then the
impurity-scattering assisted electron umklapp scattering rate in
Eq. (\ref{scattering-rate})]. This impurity-scattering assisted kernel function
$F(\theta,\theta')$ is proportional to the full effective spin propagator
$P({\bf k},{\bf p},{\bf k}',\omega)$, however, as we have mentioned in
Eq. (\ref{effective-spin-excitation}), the full effective spin excitation energy
scales with $p^{2}$, and then when the impurity-scattering assisted electron umklapp
scattering kicks in, the energy scale is proportional to $\Delta^{2}_{p}$ due to the
presence of this $p^{2}$ scaling in the full effective spin excitation energy. This
leads to that $T_{\rm scale}=\bar{b}\Delta^{2}_{p}$ can be refereed to as {\it the
temperature scale}, where the average value
$\bar{b}=(1/N)\sum\limits_{{\bf q}\in \{{\bf k}_{\rm A}\}}b({\bf q})$ is a constant
at a given doping, and the summation ${\bf q}\in \{{\bf k}_{\rm A}\}$ is restricted
to the extremely small area $\{{\bf k}_{\rm A}\}$ around the ${\bf k}_{\rm A}$ point
of the Brillouin zone. In the recent discussions \cite{Ma24}, we have shown that this
$T_{\rm scale}$ is intrinsically correlated with the antinodal spin pseudogap, and
therefore is strongly doping dependent. For a convenience in the following
discussions, $T_{\rm scale}$ as a function of doping is {\it reploted} in
Fig. \ref{T-scale-doping}, where $T_{\rm scale}$ presents clearly a similar behavior
of $T^{*}_{\rm s}$ shown in Fig. \ref{spin-pseudogap-doping}b in the underdoped
pseudogap phase, and is very low in the overdoped strange-metal phase due to the
absence of the antinodal spin pseudogap shown in Fig. \ref{spin-pseudogap-doping}a.
With the help of the above scaling of the full effective spin excitation energy in
Eq. (\ref{effective-spin-excitation}) and the doping dependence of $T_{\rm scale}$
in Fig. \ref{T-scale-doping}, our previous analysis \cite{Ma23,Ma24} has demonstrated
clearly that the impurity-scattering assisted umklapp-scattering effect is not
exponentially small at low temperatures as in the case of the electron-phonon
coupling, but is power law down to zero temperature, where three primary regions of
the impurity-scattering assisted kernel function need to be distinguished:\\
(i) in the low-temperature region ($T<T_{\rm scale}$) in the underdoped pseudogap
phase shown in Fig. \ref{T-scale-doping}, the impurity-scattering assisted kernel
function $F(\theta,\theta')$ is reduced as $F(\theta,\theta')\propto T^{2}$, this
leads to a low-temperature T-quadratic resistivity $\rho(T)\propto T^{2}$ shown in
Fig. \ref{resistivity-temperature-underdoping} in the underdoped pseudogap phase;
\\
(ii) however, in the low-temperature region ($T>T_{\rm scale}$) in the overdoped
strange-metal phase shown in Fig. \ref{T-scale-doping}, the impurity-scattering
assisted kernel function $F(\theta,\theta')$ is reduced as
$F(\theta,\theta')\propto T$, which generates a T-linear resistivity
$\rho(T)\propto T$ shown in Fig. \ref{resistivity-temperature-overdoping} in the
overdoped strange-metal phase. Concomitantly, the low-temperature resistivity
exhibits an
exceptional crossover from the T-linear behaviour in the overdoped strange-metal
phase to the T-quadratic behaviour in the underdoped pseudogap phase.\\
(iii) on the other hand, in the far-lower-temperature region
$T<T_{\rm scale}\sim 0.01J$ in the overdoped strange-metal phase, the
impurity-scattering assisted kernel function $F(\theta,\theta')$ is reduced as
$F(\theta,\theta')\propto T^{2}$, this behaviour therefore produces a T-quadratic
resistivity in the far-lower-temperature region.

\section{Summary and Discussion}\label{summary}

Starting from the microscopic electronic structure of cuprate superconductors, we
have studied the low-temperature resistivity in the normal-state from the
underdoped pseudogap phase to the overdoped strange-metal phase, and show that the
mechanism of the momentum relaxation requires both the impurity scattering and the
electron umklapp scattering. The impurity scattering is required to restrict the
modification of the distribution function to at around the antinodal region of EFS,
while the
impurity-scattering assisted electron umklapp scattering mediated by the exchange
of the full effective spin propagator dominates the temperature dependent behaviour
of the low-temperature resistivity, where the doping dependence of the temperature
scale $T_{\rm scale}$ exists, which is proportional to the square of the minimal
electron umklapp scattering vector, and presents a similar behavior of the antinodal
spin pseudogap crossover temperature. In the overdoped strange-metal phase, where
the antinodal spin pseudogap vanishes, the impurity-scattering assisted electron
umklapp scattering from a spin excitation associated with the antinodal region
produces a low-temperature T-linear resistivity in the temperature region above the
temperature scale ($T>T_{\rm scale}$), however, in the underdoped pseudogap phase,
where the antinodal spin pseudogap opens, the antinodal spin pseudogap in the
temperature region below the temperature scale ($T<T_{\rm scale}$) suppresses
partially the spin excitation density of states at around the antinodal region,
which weakens the strength of the impurity-scattering assisted electron umklapp
scattering at around the antinodal region, and thus leads to a low-temperature
T-quadratic resistivity. As a natural consequence, the low-temperature resistivity
exhibits an unusual crossover from the low-temperature T-linear behaviour in the
overdoped strange-metal phase to the low-temperature T-quadratic behaviour in the
underdoped pseudogap phase.\\
~~\\
{\bf Author Contributions}: X.M. contributed numerical calculations of the
low-temperature resistivity. Writing-original draft, S.F., X.M., M.Z., and H.G. All
authors participated in the conceptualization, theoretical research, data analysis
and interpretation, discussions, and approval of this manuscript. All authors have
read and agreed to the published version of the manuscript.\\
~~\\
{\bf Funding}: This work is supported by the National Key Research and Development
Program of China under Grant Nos. 2021YFA1401803 and 2023YFA1406500, the National
Natural Science Foundation of China under Grant Nos. 12504172, 12574249, and 12274036,
and the Special Funding for Postdoctoral Research Projects in Chongqing under Grant
No. 2024CQBSHTB3156.\\
~~\\
{\bf Acknowledgments}: The authors would like to thank Dr. X. T. Zhang for the
helpful discussions.\\
~~\\
{\bf Conflicts of Interest}: The authors declare no conflicts of interest.

\begin{appendix}

\section{Numerical Solution of Boltzmann Transport Equation}\label{NSBE}

In this Appendix, we solve numerically the Boltzmann transport equation by discretizing the
$\theta$ variable in Eq. (\ref{electron-collision}) of the main text, where the
integral-differential equation (\ref{electron-collision}) is converted to a matrix
{equation as,} 

\begin{eqnarray}\label{numerical-electron-collision}
e{\hat{\bf v}}_{\rm F}\cdot {\bf E}=-2\hat{F}\cdot\hat{\Phi},~~~~~
\end{eqnarray}
where ${\hat{\bf v}}_{\rm F}$ and $\hat{\Phi}$ can be expressed explicitly as,
\begin{eqnarray}\label{VF-Phi}
{\hat{\bf v}}_{\rm F}=
\left(
\begin{array}{c}
{\bf v}_{\rm F}(\theta_{1})\\
{\bf v}_{\rm F}(\theta_{2})\\
{\bf v}_{\rm F}(\theta_{3})\\
\cdot\\
\cdot\\
\cdot\\
{\bf v}_{\rm F}(\theta_{N})
\end{array}\right),~~~~~~~~
\hat{\Phi}=
\left(
\begin{array}{c}
\Phi(\theta_{1})\\
\Phi(\theta_{2})\\
\Phi(\theta_{3})\\
\cdot\\
\cdot\\
\cdot\\
\Phi(\theta_{N})
\end{array}\right),~~~~~
\end{eqnarray}
respectively, while the matrix $\hat{F}$ can be derived straightforwardly as,
\begin{widetext}
\begin{eqnarray}\label{F-matrix}
\hat{F}&=&
{1\over N_{\theta}}\left(
\begin{array}{ccccc}
\sum\limits_{\theta'_{l}\neq\theta_{1}}\zeta(\theta'_{l})F(\theta_{1},\theta'_{l})
+\gamma_{0}, &-\zeta(\theta'_{2})F(\theta_{1},\theta'_{2}), &\cdots,
&-\zeta(\theta'_{N})F(\theta_{1},\theta'_{N})\\
-\zeta(\theta'_{1})F(\theta_{2},\theta'_{1}),
&\sum\limits_{\theta'_{l}\neq\theta_{2}}\zeta(\theta'_{l})F(\theta_{2},\theta'_{l})
+\gamma_{0}, &\cdots, &-\zeta(\theta'_{N})F(\theta_{2},\theta'_{N})\\
\vdots &\vdots &\vdots &\vdots \\
-\zeta(\theta'_{1})F(\theta_{N},\theta'_{1}),
&-\zeta(\theta'_{2})F(\theta_{N},\theta'_{2}), &\cdots,
&\sum\limits_{\theta'_{l}\neq\theta_{N}}\zeta(\theta'_{l})F(\theta_{N},\theta'_{l})
+\gamma_{0}
\end{array}\right).
~~~~~
\end{eqnarray}
\end{widetext}
It should be emphasized that the impurity-scattering rate $\gamma_{0}$ as the diagonal
elements emerges in the above kernel function matrix $\hat{F}$, and in this sense,
this kernel function (then the electron umklapp scattering) is refereed to as the
impurity-scattering assisted kernel function (then the impurity-scattering assisted
electron umklapp scattering). In principle, the clean limit can be approached by
continuing to decrease the impurity-scattering rate $\gamma_{0}$. However, the
important electron umklapp scattering process is concentrated at around the antinodal
region \cite{Ma23,Ma24,Lee21}, where only small momentum scattering dominates the
electron umklapp scattering as we have mentioned in the main text. In this case, the
impurity scattering is required to restrict the modification of the distribution
function to the antinodal region \cite{Lee21}. This follows from a basic fact that in
the clean limit with the impurity-scattering rate $\gamma_{0}=0$, the determinant of
the above matrix $\hat{F}$ is equal to zero, i.e., $|\hat{F}|=0$, which leads to that
the region of the temperature dependent resistivity due to the electron umklapp
scattering from any a boson mode without the impurity-scattering assistance shrinks
to zero \cite{Lee21}. This is why the mechanism of the momentum relaxation from any a
boson mode requires both the impurity scattering and the electron umklapp scattering.

By numerically solving the inverse matrix $\hat{F}^{-1}$, we can obtain directly
the accurate solution for $\hat{\Phi}$ from the above matrix equation
(\ref{numerical-electron-collision}) as,
\begin{eqnarray}\label{numerical-solution-electron-collision}
\hat{\Phi}=-{e\over 2}\hat{F}^{-1}{\hat{\bf v}}_{\rm F}\cdot {\bf E}.~~~~~
\end{eqnarray}
Substituting the above result of $\hat{\Phi}$ into
Eq. (\ref{dc-conductivity}) of the main text, the dc conductivity (then the
resistivity) can be accurately evaluated.

\end{appendix}

\end{document}